\documentclass[11pt,a4paper,colorlinks]{article}
\usepackage{subfig}
\pdfoutput=1
\usepackage{longtable}
\usepackage{jcappub}
\arxivnumber{1201.4711}
\title{Indications for a pair-production anomaly from the propagation of VHE
gamma-rays}
\author[a]{D. Horns}
\author[a]{M. Meyer}

\affiliation[a]{Institute for Experimental Physics, University of Hamburg,
Luruper Chaussee 149, D-22761 Hamburg }
\emailAdd{dieter.horns@physik.uni-hamburg.de}

\abstract{In the recent years, the number of detected very high energy (VHE: $E>100$~GeV) gamma-ray sources has increased rapidly. 
The sources have been observed at redshifts up to $z=0.536$ without strong indications for the presence of absorption features
in the energy spectra. Absorption is however expected due to pair-production processes of the
propagating photons with the photon bath in intergalactic space. Even though this photon density is not well known, lower limits can be
firmly set by the resolved emission from galaxy counts. Using this guaranteed background light, we investigate the behaviour of the
energy spectra in the transition region from the optically thin to the optically thick regime. Among the sample of 50 energy spectra, 
7 spectra cover the the range from optical depth $\tau<1$ to $\tau>2$. For these sources, the 
transition to $\tau>2$ takes place at widely different energies ranging from $0.4$~TeV to $21$~TeV. Consistently, in all of these sources,
an upturn of the absorption-corrected spectrum is visible at this transition with a combined significance of $4.2$~standard deviations. 
 Given the broad range of energies and redshifts covered by the sample,  source-intrinsic features are unlikely to explain the observed effect. 
Systematic effects related to observations have been investigated and found to be not sufficient to account for the observed effect.
The pair-production process seems to be suppressed in a similar way as expected in the extension of the standard model by a light ($<$neV) 
pseudoscalar (axion-like) particle.  }
\arxivnumber{1234.5678}
\begin{document}
\maketitle
\section{Introduction}
 The pair-production process $\gamma + \gamma \rightarrow e^+ e^-$ is 
a well-understood process mostly
studied in the laboratory  through its inverse process of pair annihilation
(e.g. in storage ring experiments). 
Energetic photons propagating through the intergalactic space  undergo 
pair production with low energy photons  of the extragalactic background light (EBL) in the optical
 (mostly star-light) and infrared (re-emitted light from warm and cold dust)
which  leads to the attenuation of the primary beam 
(e.g. \cite{1967PhRv..155.1404G}) as well as to the formation of an
inverse-Compton/pair-production cascade (e.g. \cite{1986MNRAS.221..769P}).
Secondary  emission from these cascades may even dominate 
the observed emission (see e.g. \cite{2011PhRvD..84h5019A,2011arXiv1107.5576M} for gamma-ray induced cascades and 
\cite{2006PhRvD..73h3008A,2010APh....33...81E,2011ApJ...731...51E} for ultra-high energy proton induced cascades)\footnote{the observed
broad-band gamma-ray variability of distant Blazars as e.g. H1426+428 ($z=0.129$) \cite{2003A&A...403..523A,2011arXiv1110.0038W},  1ES0229+200 ($z=0.140$) 
\cite{2011arXiv1110.0038W}, and 1ES1218+304 ($z=0.184$) \cite{ASIF_IMRAN} 
 indicates however that the bulk of the observed emission is very likely not produced in cascades.}.
 \\
In the context of propagation of unpolarized energetic photons, a number of 
processes have been suggested to modify the standard model behaviour:
The pair production process could be 
affected by Lorentz-invariance violation (LIV) processes \cite{1999PhRvD..59k6008C,2008PhRvD..78l4010J}, 
kinematic mixing with hidden sector photons \cite{2008JHEP...07..124A}, 
as well as conversion
and re-conversion of photons into axion-like particles (ALPS) \cite{2003JCAP...05..005C,2007PhRvD..76l1301D}. 
 Such effects modify
the resulting optical depth (increasing as well as decreasing it)  and 
 may even depend on the particular line of
sight to the source \cite{2009JCAP...12..004M}.  Therefore, careful spectroscopy of
gamma-ray sources in the optically thick  regime could effectively
probe the existence of pair-production anomalies.
\\ 
The effect of pair-production during propagation should lead to a pronounced softening of the
observed spectra from extragalactic sources with increasing optical depth. 
 Remarkably, a systematic softening of the sources with e.g.
increasing redshift has not been observed so far \cite{2011arXiv1106.1132D}.  This fact itself is
surprising, but may be explained by e.g.  reducing the assumed density of
absorbing photons or more subtly, observational effects without invoking any
anomalies. \\
Conversely, the apparent lack of absorption has been
used to constrain the level of extragalactic background light (EBL) by
assuming that the intrinsic shape of the source spectrum should not be harder
than plausible models suggest. 
These analyses 
constrain the maximum level of the EBL  consistently to be close to 
\cite{2006Natur.440.1018A,2005ApJ...618..657D, 2007A&A...471..439M} 
and even slightly below (in the mid-infrared) \cite{2011ApJ...733...77O} the 
guaranteed level of the EBL \cite{2010A&A...515A..19K}. 
The observations of the spectra at
energies where the optical depth is low (e.g. with Fermi/LAT) or observation of
nearby sources provides a check on the assumptions for the source spectra in an unbiassed way 
\cite{2011ApJ...733...77O}.  \\
So far, attempts to search for deviations from the expected optical depth have mostly relied
on the examination of the power law index of the observed gamma-ray spectrum
as a function of redshift. The absence of  systematical
softening has been interpreted as an indication for ALPS-conversion
processes \cite{2011arXiv1106.1132D,2009PhRvD..79l3511S}. A more sophisticated approach was
followed in \cite{2011JCAP...11..020D} where the measured spectra of 3C279, 3C 66A, PKS~1222+216, and
PG~1553+113 were fit taking the modification of ALPS-related effects into account.
In the following, we search for systematic effects in the energy spectra at the
expected transition from optically thin to optically thick 
in the gamma-ray energy spectra.   
\section{Data analysis and results}
 \subsection{Summary of data used}
\begin{figure}
\begin{center}
 \includegraphics[width=0.8\linewidth]{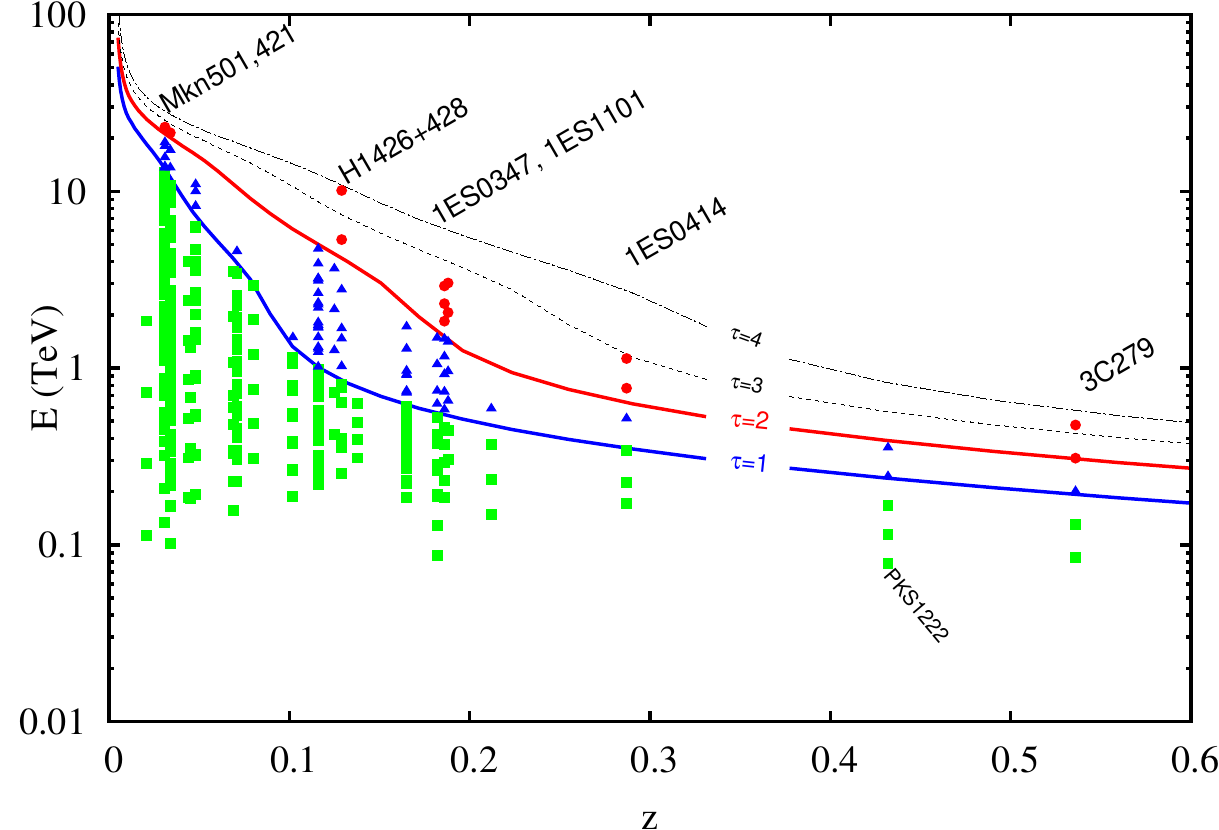}
 \caption{\label{fig:over} A summary of the spectral data used in this study: 
For each individual spectral measurement, the corresponding value of $z$ and $E$ are marked in this diagram. 
Overlaid are the iso-contours for $\tau=1,2,3,4$ calculated using the minimum EBL model.}
\end{center}
\end{figure}

  We have extracted from the literature all available individual measurements
of the differential flux (spectral points) from extragalactic sources with
known redshifts\footnote{nearby sources like M87 and Cen A have been excluded}. 
In total, 389 individual spectral points from 50 spectral measurements of 25 sources have
been accumulated.   A summary of the data is
given in Table~\ref{table:data} including
references. For a number of sources, the spectra have
been measured at different times and with different instruments. Statistically independent 
measurements of identical sources are included in the sample. 
Objects without confirmed spectral determination of the  redshift are excluded from the sample 
as well
as individual measurements which at a later stage have been re-analysed or combined
in a time-average measurement. \\
The observed spectral points $\varphi_i(E_i)$ for each source at redshift $z$  are assigned 
an optical depth $\tau_i(E_i,z)$ using the minimal EBL model \cite{2010A&A...515A..19K}. In Fig.~\ref{fig:over}, the
locations of the measurements in the $z,E$-plane are indicated by an individual marker. Overlaid are the lines for constant
optical depth $\tau=1,2,3,4$. 
 For each observed spectral point, we can then readily
calculate the spectral point corrected for the effect of absorption
\begin{eqnarray}
\Phi_i=\exp(\tau_i) \varphi_i. 
\end{eqnarray}
\begin{longtable}{llccc|p{2.2cm}}
\hline
\hline
id & Source & Redshift & Experiment & Energy Range &  Reference\\
   &         &          &            & (TeV)        &           \\
\hline
\endfirsthead
\multicolumn{6}{c}{
  {\bfseries \tablename \thetable -- continued from previous page}} \\
\hline
\hline
id   & Source & Redshift & Experiment & Energy Range &  Reference\\
     &        &          &            & (TeV)        &           \\
\hline
\endhead

\hline
\multicolumn{6}{c}{
 {Continued on next page} }\\ \hline
\endfoot
\hline\hline
\caption{TeV blazar spectra used in this paper ordered by red-shift. The spectra which contain data-points with
optical depth $\tau>2$ are marked with bold-face.\label{table:data} }
\endlastfoot
 1&3C66B & 0.021 & MAGIC & 0.11 -- 1.85  & {\cite{2009ApJ...692L..29A}}\\		
 2&Mkn421 & 0.031 & HEGRA & 0.56 -- 6.86  & {\cite{1999A&A...350..757A}}\\  
3&Mkn421 & 0.031 & HEGRA & 0.82 -- 13.59  & {\cite{2002A&A...393...89A}}\\ 
4&Mkn421 & 0.031 & HEGRA & 0.82 -- 13.59  & {\cite{2002A&A...393...89A}}\\ 
5&Mkn421 & 0.031 & HESS  & 1.12 -- 17.44  & {\cite{2005A&A...437...95A}}\\ 
6&Mkn421 & 0.031 & WHIPPLE & 0.38 -- 8.23  & {\cite{2002ApJ...575L...9K}}\\
7&Mkn421 & 0.031 & MAGIC & 0.13 -- 1.84  & {\cite{2007ApJ...663..125A}}\\  
8&Mkn421 & 0.031 & MAGIC & 0.45 -- 4.24  & {\cite{2010A&A...519A..32A}}\\  
\bf 9&\bf Mkn421 & \bf 0.031 & \bf HESS  &\bf  1.73 -- 23.1  & {\cite{2011arxiv1106.1035T}}\\  
\bf10 &\bf Mkn501 &\bf  0.034 &\bf  HEGRA &\bf  0.56 -- 21.45  & {\cite{1999A&A...349...11A}}\\ 
11&Mkn501 & 0.034 & CAT & 0.40 -- 10.00  & {\cite{1999A&A...350...17D}}\\  	
12&Mkn501 & 0.034 & VERITAS & 0.27 -- 3.86  & {\cite{2009arXiv0912.3772H}}\\ 
13&Mkn501 & 0.034 & VERITAS & 0.22 -- 1.90  & {\cite{2009arXiv0912.4728G}}\\
14&Mkn501 & 0.034 & VERITAS & 0.25 -- 3.89  & {\cite{2011ApJ...727..129A}}\\ 
15&Mkn501 & 0.034 & MAGIC & 0.17 -- 4.43  & {\cite{2011ApJ...727..129A}}\\  
16&Mkn501 & 0.034 & VERITAS & 0.26 -- 3.80  & {\cite{2009arXiv0912.3772H}}\\ 
17&Mkn501 & 0.034 & MAGIC & 0.10 -- 1.76  & {\cite{2009ApJ...705.1624A}}\\  
18&Mkn501 & 0.034 & VERITAS & 0.25 -- 3.81  & {\cite{2011ApJ...727..129A}}\\ 
19&1ES2344+514 & 0.044 & MAGIC & 0.19 -- 4.00  & {\cite{2007ApJ...662..892A}}\\   
20&Mkn180      & 0.045 & MAGIC & 0.18 -- 1.31  & {\cite{2006ApJ...648L.105A}}\\  
21&1ES1959+650 & 0.048 & HEGRA & 1.59 -- 10.00  & {\cite{2003A&A...406L...9A}}\\  
22&1ES1959+650 & 0.048 & HEGRA & 1.52 -- 10.94  & {\cite{2003A&A...406L...9A}}\\ 
23&1ES1959+650 & 0.048 & MAGIC & 0.19 -- 1.53  & {\cite{2006ApJ...639..761A}}\\ 
24&1ES1959+650 & 0.048 & MAGIC & 0.19 -- 2.40  & {\cite{2008ApJ...679.1029T}}\\  
25&BLLacertae & 0.069 & MAGIC & 0.16 -- 0.70  & {\cite{2007ApJ...666L..17A}}\\ 
26&PKS0548-322 & 0.069 & HESS & 0.34 -- 3.52  & {\cite{2010arXiv1006.5289H}}\\ 
27&PKS2005-489 & 0.071 & HESS & 0.23 -- 2.27  & {\cite{2005A&A...436L..17A}}\\ 
28&PKS2005-489 & 0.071 & HESS & 0.34 -- 4.57  & {\cite{2010A&A...511A..52H}}\\ 
29&RGBJ0152+017 & 0.080 & HESS & 0.31 -- 2.95  & {\cite{2008A&A...481L.103A}}\\ 
30&W Comae & 0.102 & VERITAS & 0.26 -- 1.15  & {\cite{2008ApJ...684L..73A}}\\ 
31&W Comae & 0.102 & VERITAS & 0.19 -- 1.49  & {\cite{2009ApJ...707..612A}}\\ 
32&PKS2155-304 & 0.116 & HESS & 0.23 -- 2.28  & {\cite{2005A&A...430..865A}}\\
33&PKS2155-304 & 0.116 & HESS & 0.23 -- 3.11  & {\cite{2005A&A...442..895A}}\\
34&PKS2155-304 & 0.116 & HESS & 0.22 -- 4.72  & {\cite{2007ApJ...664L..71A}}\\
35&PKS2155-304 & 0.116 & HESS & 0.25 -- 3.20  & {\cite{2009ApJ...696L.150A}}\\
36&RGBJ0710+591 & 0.125 & VERITAS & 0.42 -- 3.65  & {\cite{2010ApJ...715L..49A}}\\
\bf 37&\bf H1426+428 &\bf  0.129 &\bf  HEGRA,CAT,WHIPPLE &\bf  0.25 -- 10.12  & {\cite{2003A&A...403..523A}}\\
38&1ES0806+524 & 0.138 & MAGIC & 0.31 -- 0.63  & {\cite{2009ApJ...690L.126A}}\\
39&1ES0229+200 & 0.140 & HESS & 0.60 -- 11.45  & {\cite{2007A&A...475L...9A}}\\
40&H2356-309 & 0.165 & HESS & 0.22 -- 0.91  & {\cite{2006A&A...455..461A}}\\
41&H2356-309 & 0.165 & HESS & 0.23 -- 1.71  & {\cite{2010A&A...516A..56H}}\\
42&H2356-309 & 0.165 & HESS & 0.18 -- 0.92  & {\cite{2006Natur.440.1018A}}\\
43&1ES1218+304 & 0.182 & MAGIC & 0.09 -- 0.63  & {\cite{2006ApJ...642L.119A}}\\
44&1ES1218+304 & 0.182 & VERITAS & 0.19 -- 1.48  & {\cite{2009ApJ...695.1370A}}\\
\bf 45&\bf 1ES1101-232 &\bf  0.186 &\bf  HESS &\bf  0.18 -- 2.92  & {\cite{2006Natur.440.1018A}}\\
\bf 46& \bf  1ES0347-121 & \bf 0.188 & \bf HESS & \bf 0.30 -- 3.03  & {\cite{2007A&A...473L..25A}}\\
47&1ES1011+496 & 0.212 & MAGIC & 0.15 -- 0.59  & {\cite{2007ApJ...667L..21A}}\\
\bf 48&\bf 1ES0414+009 & \bf 0.287 & \bf HESS &  \bf 0.17 -- 1.13  & \cite{2012arXiv1201.2044T} \\
49&PKS1222+21 & 0.432 & MAGIC & 0.08 -- 0.35  & {\cite{2011arXiv1101.4645A}}\\
\bf 50&\bf 3C279 & \bf 0.536 &\bf    MAGIC &\bf  0.08 -- 0.48  & {\cite{2008Sci...320.1752M}}\\
\end{longtable}

 \subsection{Method to search for the anomaly}
 For each spectrum, the data points observed in the optically thin regime are identified by requiring $\tau<1$. These data points
are the basis to determine the parameters of a fitting function $f_\mathrm{id} (E)$ to  $\Phi_i$. The fitting function 
describes the shape of the energy spectrum within 
the optically thin regime and therefore should be close to the intrinsic spectral shape. 
The fitting function is a power law of the form $f_{id}(E)=f_0 (E/E_d)^{-\Gamma}$ with two free parameters for each source 
($\mathrm{id}$): the
normalization $f_0$ and the photon index $\Gamma$. The best-fitting  parameters are found with a $\chi^2$-minimization procedure. The
decorrelation energy $E_d$ is chosen such that the covariance matrix is diagonal. The $p$-value for the resulting value of $\chi^2$ and degrees-of-freedom
($dof$) of  the fit is calculated and used
as a goodness-of-fit estimator. If the $p$-value is smaller than $0.05$, a more complex fitting function is chosen 
to be a log-log parabola $f_\mathrm{id}(E)=f_0 (E/E_d)^{-\Gamma + \beta \log(E/E_d)}$.  The curved spectrum is used to fit the
energy spectra of Mkn~421 ($id=2,3,5,7$), Mkn~501 ($id=10,11$), and PKS~2155-305 ($id=34$) 
during high flux states satisfactorily. Any spectra which can not be
described with $p>0.05$ ($id=7$) or have less than two spectral points with $\tau<1$ ($id=39$) are rejected from the sample. The final data sample
consists of 389 spectral points with 305 optically thin measurements.

The data points with optical depth $\tau\ge 1$ are further split into a reference (base) sample $\mathcal{B}=\{\Phi_i|1\le \tau_i<2\}$ 
and the search sample $\mathcal{S}=\{\Phi_i|2\le \tau_i\}$.
The intervals chosen are guided by the expected effect of the coupling to an axion-like particle which leads to a boost of the
observed flux at optical depth beyond approximately two \cite{2009PhRvD..79l3511S}. 

\begin{figure}
 \begin{center}
 \includegraphics[width=0.8\linewidth]{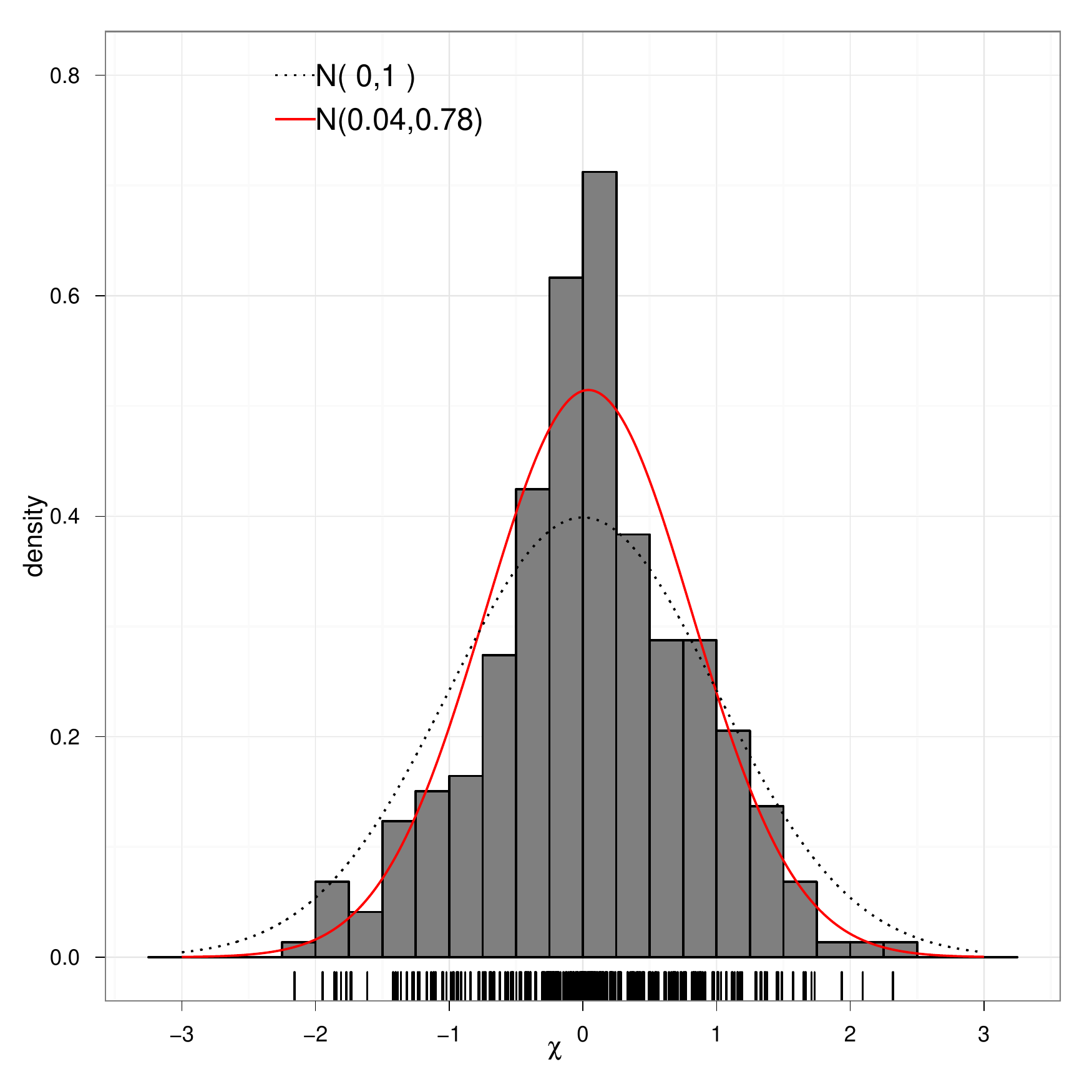}
 \caption{\label{fig1} The distribution of the error-normalized scatter of individual values of $\Phi_i$ which enter the
fit ($\tau<1$): For comparison,
a normal distribution $N(0,1)$ is overlaid (dashed line) together with the best fitting normal distribution with mean $0.04\pm0.05$ and
width $\sigma=0.78\pm0.03$.}
 \end{center}
\end{figure}

For each flux point in $\mathcal{B}$ and $\mathcal{S}$, we calculate a quantity which provides a measure on how the flux points scatter
around the extrapolated expectation from the optically thin spectrum:
\begin{eqnarray}
 	R(\Phi_i) = \frac{\Phi_i - f_\mathrm{id}(E_i) }{\Phi_i + f_\mathrm{id}(E_i)}.
\end{eqnarray}
 So far, the estimated uncertainties (both statistical and systematical) on
$\varphi_i$ have been ignored. The systematic effects will be subject of
discussion in the following subsection (\ref{subsec:systematics}). The
statistical uncertainties can not be included in the test, because the observed
scatter of the 
measurements around the fit indicate that the error estimates are larger than the actual scatter. This follows from the consideration of the distribution of 
the normalized residuals
\begin{eqnarray}
 \chi_i &:=& \frac{\Phi_i-f_\mathrm{id}(E_i)}{\sigma(\Phi_i)},
\end{eqnarray} 
which does not follow a $N(\mu,\sigma)$ normal distribution with 
$\mu=0$ and $\sigma=1$ as expected. An unbinned
likelihood fit of $\chi$ provides best-fit estimators for 
$\mu=0.04\pm0.05$ and $\sigma=0.78\pm0.03$ (errors quoted are estimated for $68~\%$ confidence intervals). 
Interestingly, the distribution is barely compatible with a Gaussian (performing an Anderson-Darling test, 
the probability for a normal distribution is $1.7~\%$). As can be seen in Fig.~\ref{fig1}, the distribution shows
tails towards both negative as well as positive values indicating that the power law assumption is in slight tension with the actual spectra. 
Similar results for the width of the distribution of $\chi$ have been obtained for spectra of Galactic sources.
  It remains unclear why the
true observational uncertainties are universally smaller than the estimated ones. In principle, the errors could
be scaled by a factor $0.78$ in order to match the errors with the observed scatter, but it appears difficult to draw firm conclusions on
scaled errors. {A close inspection of the residuals
in different intervals of energy and optical depth (see Appendix B) indicates
that the scatter of the residuals varies for the different samples
 considered,
making it unreasonable to apply a global scaling of the estimated uncertainties.}
Instead, the test advocated here takes 
self-consistently the scatter of the data into account and does not rely on properly estimated errors. 
\subsection{Results of the test}
\begin{figure}
 \begin{center}
 \includegraphics[width=0.8\linewidth]{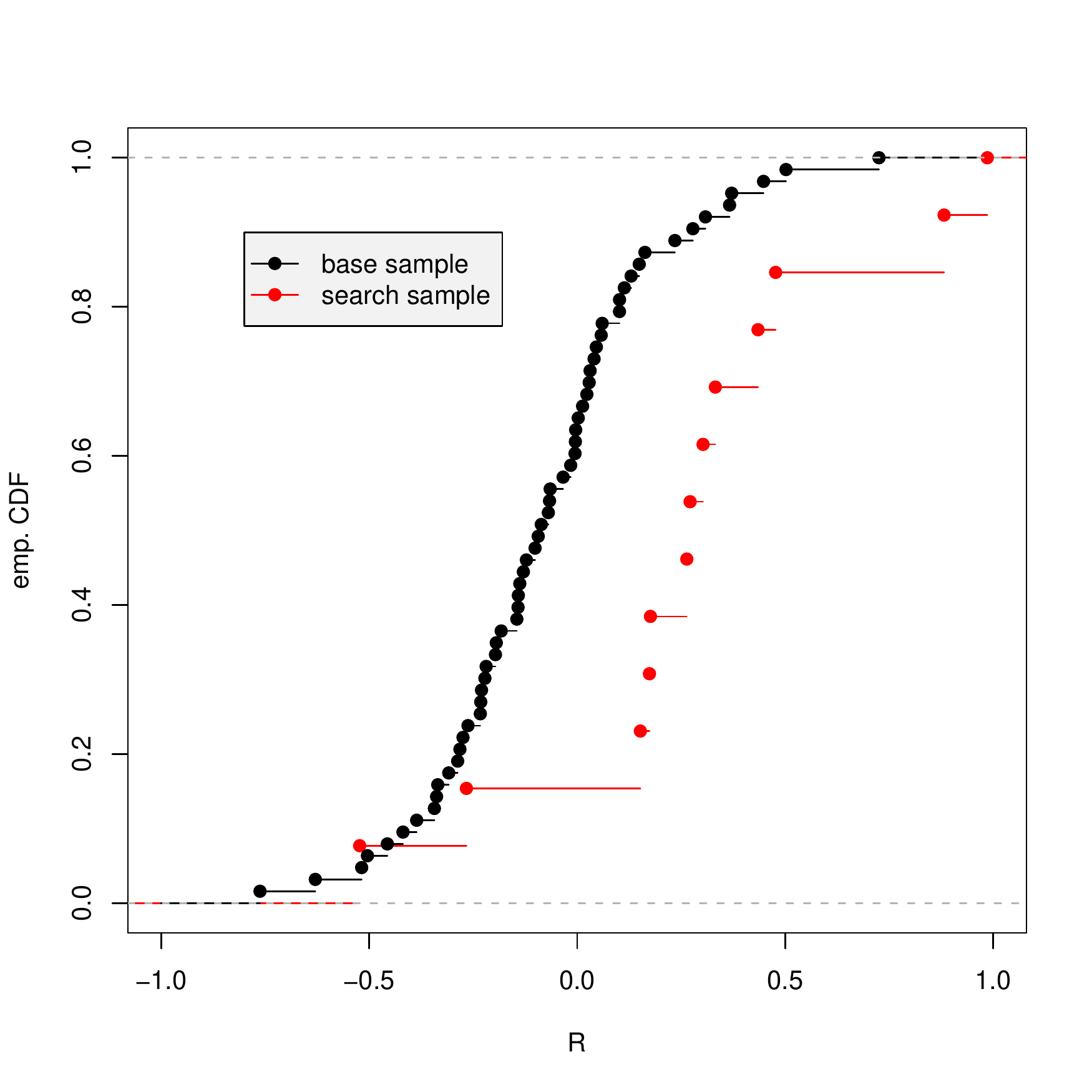}
  \caption{\label{fig2} For the two samples ($\mathcal{B}$: base and $\mathcal{S}$: search) the empirical cumulative distribution functions (CDF) 
are 
compared. The two-sample Kolmogorov-Smirnov test statistics indicate that the probability for the two samples to be drawn from the same underlying
distribution is $1.7\times 10^{-5}$ (corresponding to $S=4.2~\sigma$).} 
 \end{center}
\end{figure}

\begin{figure}
 \includegraphics[width=0.8\linewidth]{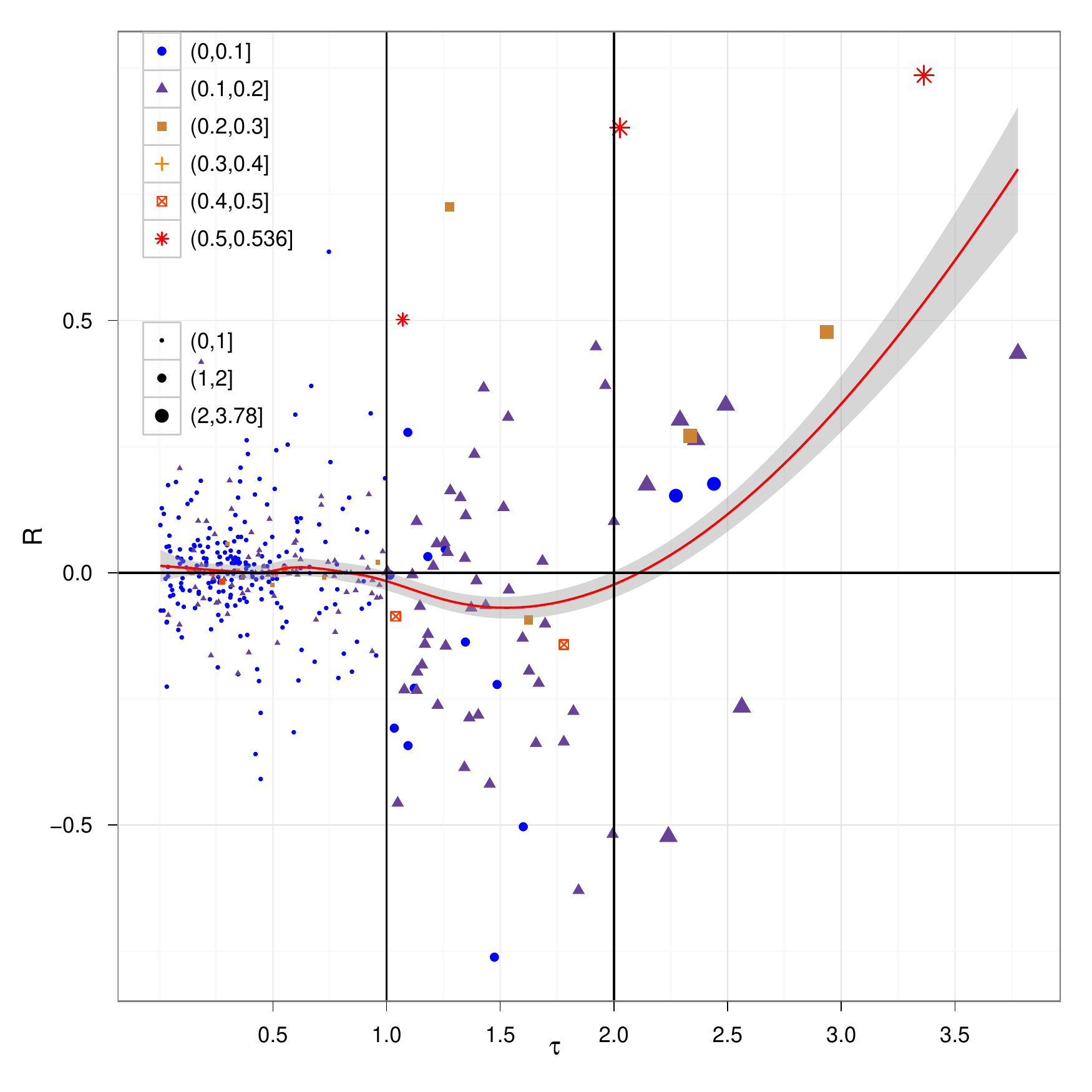}
 \caption{\label{fig3} Scatter-plot of $R$ vs $\tau$. The color/shape of the markers indicate the redshift interval and the size of the marker is proportional to the optical depth. 
The solid line and grey band are derived from a non-parametric linear smoothing method (loess) and an estimate of the corresponding confidence interval (95~\% c.l.)
\cite{loess}.} 
\end{figure}

The distribution of $R(\Phi_i\in \mathcal{B})$ with $N_1=63$ values  
is finally compared with the distribution of $R(\Phi_i \in \mathcal{S})$  ($N_2=13$) 
using the unbinned Kolmogorov-Smirnov (K-S) test on the empirical cumulative distribution function (CDF) (the individual spectra which contributed to the
$\mathcal{S}$-sample  are shown in Fig.~\ref{fig:spectra}). 
The K-S test does not rely on error estimates and is mainly sensitive to a relative shift of the two
distributions while the presence of tails or a difference in the distribution widths does not strongly affect the test.
 For the maximum difference of the two CDFs of $D=0.703$, the resulting probability that the two distributions
originate from the same parent distribution is $p(>D,N_1,N_2)=1.7\times 10^{-5}$ corresponding to a one-sided tail of a Gaussian at $S=4.2~\sigma$
(see Fig.~\ref{fig2} for a graphical representation of the CDFs.
The result indicates a shift of the distribution for the values in the $\mathcal{S}$-sample when comparing  with the $\mathcal{B}$-sample. 
The two samples
show also a quite different behaviour when considering the correlation of $R$ and $\tau$. For the sample $\mathcal{B}$, a Pearson's test on
the correlation results in $cor(\mathcal{B})=-0.09\pm0.12$ with a probability for the hypothesis of uncorrelated  data $p(\mathcal{B})=0.46$. 
The search sample shows a moderate indication for a correlation: $cor(\mathcal{S})=0.35\pm 0.25$ and $p(\mathcal{S})=0.23$. This behaviour
is apparent when producing a scatter-plot of $R$ versus $\tau$ as shown in Fig.~\ref{fig3}.
{In addition to the K-S test applied to statistically independent samples of  $R$-values, we have
used in Appendix B the sample of residuals of a fit applied to all measurements.  
 Consistent with the result of the K-S test, the average of the residuals for the sample $\chi_i(\mathcal{S})$
is significantly shifted away from zero ($\mu=0.73\pm0.13$) which indicates 
a shift of $5.6~\sigma$. It should be noted, that this result
is obtained with the additional assumption that the residuals are normal
distributed. Given the reasonable $p$-values of the Anderson-Darling test applied to the distribtion of $\chi_i$
(see Appendix B for details),
this assumption seems well justified.}

 \section{Systematic effects}
\label{subsec:systematics}
The result obtained in the previous section can be subject to systematic
effects which are not related to the propagation of the photons. In the
following, various systematic effects are considered.
\subsection{Systematic effects related to the sources}
\begin{figure}
 \includegraphics[width=0.8\linewidth]{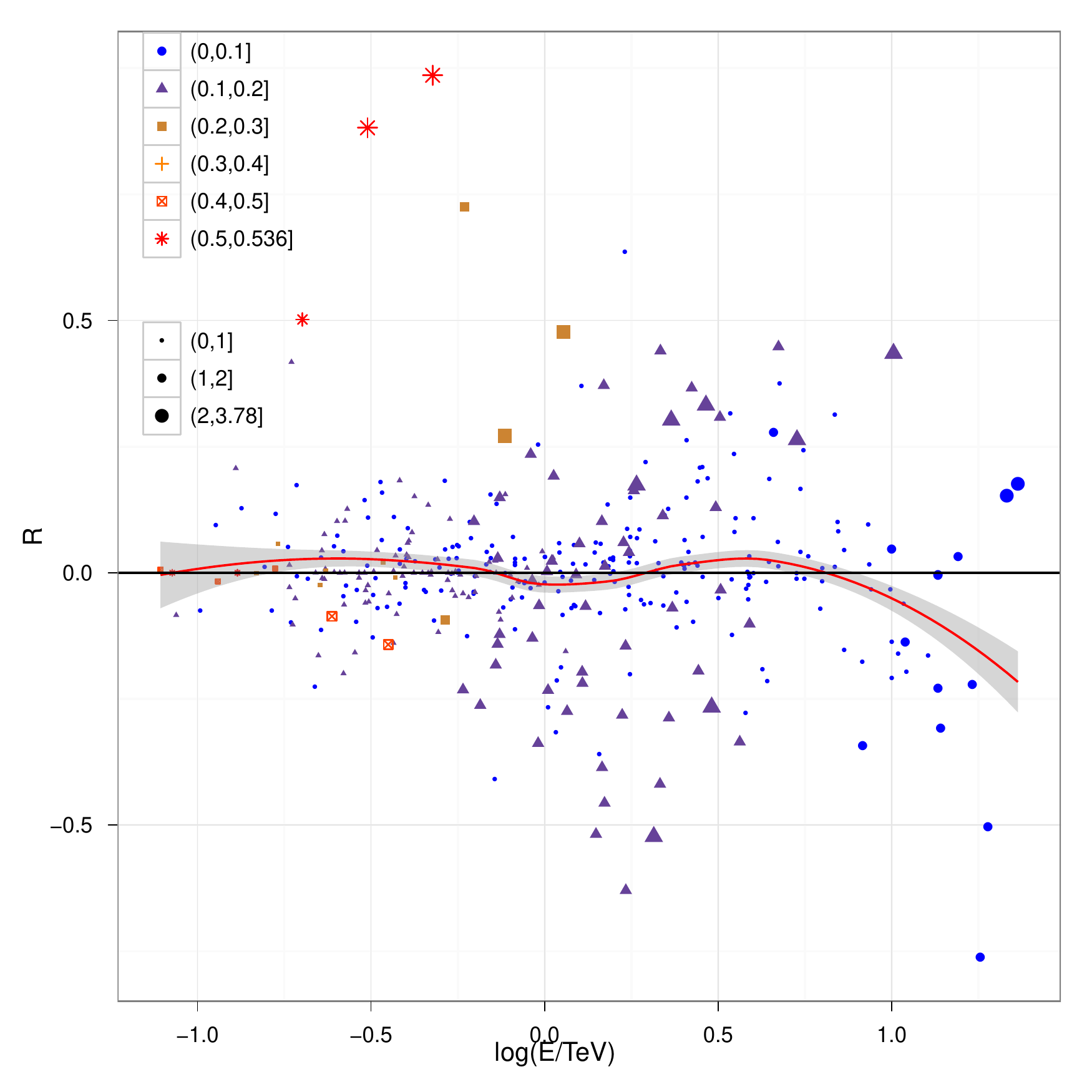}
 \caption{\label{fig4} {Scatter-plot of $R$ vs $\log_{10}(E/\mathrm{TeV})$.} The color/shape of the markers indicate the redshift interval
and the size of the marker is proportional to the optical depth. 
The solid line and grey band are derived from a quadratic smoothing method and an estimate of the corresponding confidence interval (68~\% c.l.)
\cite{loess}.} 
\end{figure}

 The choice of blazars included in the sample is strongly biased by their
detection at 100~GeV to TeV energies thus  selecting preferentially objects
with a sufficiently large luminosity. {The seven sources with measurements extending to $\tau>2$
contribute about equally to the overall excess.  Removing each source spectrum and re-calculating
the significance given in the following list provides an estimate on the relevance of each source: Mkn~421 
($S(\overline{\mathrm{Mkn~421}})=3.8~\sigma$), 
Mkn~501 ($S(\overline{\mathrm{Mkn~501}})=3.7~\sigma$),
H~1426+428 ($S(\overline{\mathrm{H~1426+428}})=3.5~\sigma$),
1ES1101-232 ($S(\overline{\mathrm{1ES1101-232}})=2.8~\sigma$),
1ES0347-121 ($S(\overline{\mathrm{1ES0347-121}})=4.6~\sigma$),
1ES0414+009 ($S(\overline{\mathrm{1ES0414+009}})=3.6~\sigma$),
3C279 ($S(\overline{\mathrm{3C279}})=3.8~\sigma$). The result demonstrates that the different objects (6 high 
frequency peaked Blazars and 1 flat spectrum radio quasar) at widely different redshifts ($z=0.031$, 
$0.034$, $0.129$, $0.186$, $0.188$, $0.287$, and $0.536$) contribute about equally to the overall significance. 
A systematic effect based upon the type of source or its distance is not evident. 
Note, the significance increases after excluding the spectrum of 1ES0347-121 from the overall test. This is a
consequence of the extra-polation of the fit to the two points at small optical depth (see Fig.~\ref{sf05}). 
}
 An unbiased or even complete catalogue of 
VHE-emitting blazars is at this point not available and will require an all-sky instrument with sufficient sensitivity (e.g. HAWC 
\cite{2011arXiv1108.6034H}).
Instead, search strategies using narrow field of view instruments rely on external triggers indicating a high-state 
of an object or selection of bright objects from radio or X-ray catalogues or a mixture of both approaches. A successful detection 
is possible if the source is sufficiently bright in gamma-rays and if the observable spectrum favors a detection. It is therefore conceivable, that
redshift dependent selection effects favor e.g. the detection of blazars with softer spectra at large redshift. Generally, it is difficult to 
predict the selection bias without further knowledge on the relation between the spectral state and the luminosity of the source and 
the analysis carried out.  \\
The test applied here does not depend on any prior assumption on the shape of the source spectrum and is therefore not sensitive to selection
biases or even multiple-components in the {high-energy part of the}
spectra
{as e.g. possibly present in
 Fermi-LAT spectra \cite{2010ApJ...716...30A}
and explained in the framework of time-dependent models for gamma-ray flares
\cite{2011ApJ...740...64L}.}
 By construction, the test only probes the transition region from $\tau<2$ to $\tau>2$. Given that
the energy at which this transition takes place varies in a non-linear way with redshift, it is very unlikely that any of the selection biases or
spectral features present in the source could depend on redshift in the 
same way. The fact that widely different sources at different redshifts
contribute to the observed anomaly strengthens this argument. \\
In Fig.~\ref{fig4}, the scatter-plot of $R$ vs $\log_{10}(E/\mathrm{TeV})$ demonstrates that the source spectra behave rather similarly across the
entire energy band covered by observations. {The residuals analysed in Appendix B show a consistent behaviour}. The notable deviations are measurements with optical depth $\tau>2$. 
 This indicates that the sources do not show any particular deviations from the assumed fitting function and its extrapolation
across the entire energy range except the optically deep regime. In conclusion, a source intrinsic hardening of the energy spectrum is not excluded
and may also be motivated in particularly tuned models \cite{2011ApJ...743L..19L}. However, there are no indications for spectral hardening in the gamma-ray spectra except for
the spectra observed at optical depth $\tau>2$ which rules out that this hardening is source-intrinsic.

 \subsection{Energy calibration}
The energy scale of the ground based Cherenkov telescopes could be erroneous.  Nominally, the experimental groups estimate the systematic uncertainty on the global
energy scale to be $\approx \pm 15~\%$ on a relative scale. In a recent study \cite{2010A&A...523A...2M}, 
the flux normalization of the Crab nebula between the different Cherenkov telescopes has been compared
among each other in order to estimate systematic differences between the energy calibration of HESS, MAGIC, and HEGRA. Relative to the lowest energy calibration (HESS),
the energy scale of MAGIC and HEGRA is shifted upwards by ($7\pm1)$~\% and ($8\pm1$)~\% respectively.  Moreover, a cross calibration with the Fermi/LAT 
demonstrates a surprisingly small difference ($<5~\%$) of the energy calibration of air shower measurements with respect to the beam-calibrated 
pair telescope. Varying the energy calibration of all instruments between 5 and 10~\% downwards leads to a maximum reduction of resulting $p$-value
to $4\times 10^{-4}$ ($3.3~\sigma$). 
For the maximum admitted shift of the energy scale ($-15$~\%), the energy spectrum of 1ES0229+200 is included in the test
leading to a smaller value of $p=2.9\times10^{-4}$ ($3.4~\sigma$). 
Nominally, this source is not included because only one data point with $\tau<1$ is not sufficient to provide a power law fit.
 After shifting the energy scale
downwards, this source is included compensating partially for the effect of the shift on other sources. 
\subsection{End-point of the energy spectra}
 The optically thick data points are naturally at the endpoint of the measured energy spectra. In this regime, the observed spectra suffer from
the limited energy resolution leading to spill-over effects. In combination with a softening of the observed energy spectrum at high energies, 
the spill-over effect could lead to a systematic over-estimate of the true flux. In combination with the exponential factor multiplied
to compensate the effect of absorption, this systematic effect could mimic the observed hardening. 
Furthermore, the endpoint of the energy spectrum is defined by the last
significant detection.  This leads to a unavoidable bias in the reconstructed flux which
tends to be larger than the true flux. Without in-depth knowledge of the
analysis methods, it is 
unrealistic to attempt to correct for these effects in a reliable way. As a simple and robust check on the bias on the analysis, 
we ignore the respective endpoint and
shift the energy scale downwards by $15$~\%, maximizing the systematic influence on the test. The resulting significance is reduced, leading
to $p=4.8\times10^{-3}$ ($S=2.6~\sigma$) which is certainly an upper limit to the true probability. 
\subsection{Mock data set}
The test carried out could be biased intrinsically by the choice of the reference and search sample. The different scatter of
the points in the two sample could lead to differences in the resulting distributions of $R$ simply by the construction (e.g. extrapolation
of the fit). 
As a mock sample, we choose energy spectra from Galactic sources with similar numbers of data points in a reference and search sample. 
Each spectrum is randomly assigned a redshift to split the data points accordingly -- however without applying 
a correction factor $\exp(\tau)$. After repeating the analysis, the resulting 
$p$-values are close to unity, demonstrating that the choice of reference and search sample does not lead automatically to differences in the
resulting distribution. 
\section{Interpretation and discussion}
 After considering and rejecting the various possibilities to
explain the observed effect by systematic uncertainties related to the
observation and the source, it appears that the most likely explanation is
related to the propagation of the photons. While a definite answer will
certainly require
a better characterization of the effect through either astrophysical observations or laboratory experiments or a combination of both, 
a number of possibilities can readily be excluded.

 \subsection{EBL}
 The level of EBL used for the test can be safely considered as a lower limit
to the actual photon density in intergalactic space \cite{2010A&A...515A..19K}.
The anomaly appears at different energies (from 360 GeV to 22 TeV) which in
turn relates to a broad range of  wavelengths of the EBL between the optical to mid-IR. 
In order to eradicate the effect entirely, the EBL (maintaining the same shape) would have to be corrected downward by about $20~\%$.
{This would reduce the number of data points in the search sample to zero.
Shifting the EBL downwards is, however, in contradiction with the observationally 
resolved part of the EBL: The scaled model falls below the lower limits derived from 
galaxy number counts in the UV / optical \cite{2000MNRAS.312L...9M} and in the NIR \cite{2004ApJS..154...39F} 
by more than one sigma of the measurement uncertainties.
Another possibility would be to shift the EBL upwards and altering its shape.
Three out of the seven sources in the sample are located at a redshift between 
$0.1 < z < 0.2$ and are measured up to a few TeV.
Thus, the absorption corrected spectra of these sources are influenced most by changes of the EBL density 
between optical and NIR wavelengths. 
Increasing the EBL density at optical wavelengths leads to softening of the EBL corrected spectra and the signal significance is reduced.
However, the effect observed for the other sources would be enhanced, partially compensating this effect.}

 \subsection{Implications Pair-production anomaly}
\begin{figure}
 \begin{center}
  \includegraphics[width=0.8\linewidth]{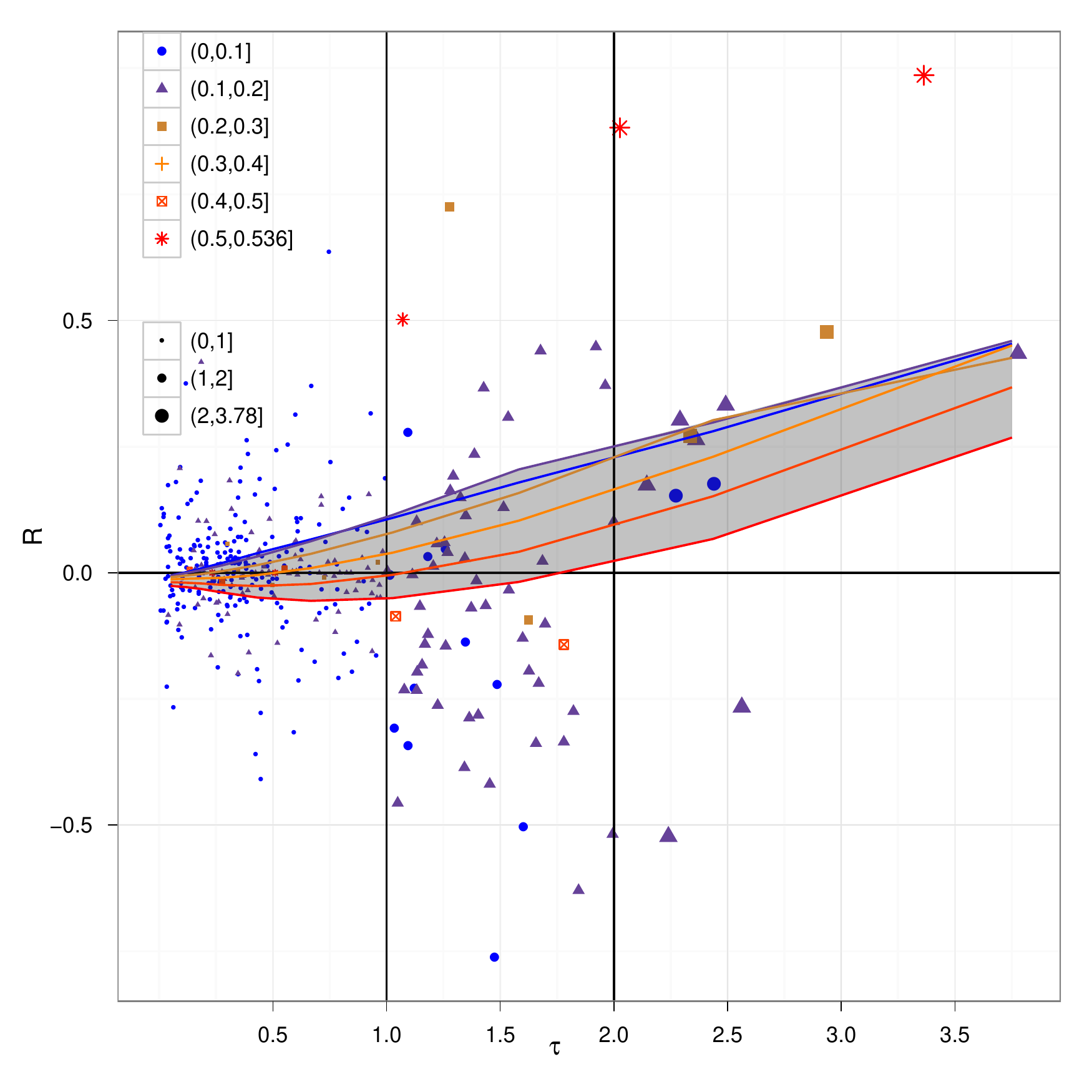}
 \caption{\label{figtau2} Overlaid on the same data as shown in Fig.~\ref{fig4} is a set of curves for different redshifts indicating
the change of the transparency if a spin-0 axion-like particle would couple to photons.
The parameters chosen were $g_{a\gamma}=10^{-11}~\mathrm{GeV}^{-1}$, $l=5$~Mpc, $B=1$~nG.}
 \end{center}
\end{figure}

After excluding the sources, the EBL, and observational effects to explain 
{all of} the 
spectral signature, we consider the possibility of 
a pair-production anomaly (PPA).  The effect could be explained if pair production in collisions
of energetic photons (from 0.3 to 20 TeV) with low energy photons of the EBL is suppressed. A number of suggestions have been discussed in the literature
which could lead to an effective pair-production anomaly. 
\begin{itemize} 
\item \textbf{Violation of Lorentz invariance:} Among the physically motivated possibilities, the violation of Lorentz invariance
and its effect on the propagation of neutral particles has been considered in
some detail. A high energy theory could in principle lead to a
modification of the photon dispersion relation which would lead to an
energy-dependent time-of-flight for photons as well as a shift of the threshold for pair production.  In principle, the shift of threshold could
lead to a suppression of pair production for higher energies. However, the observations indicate that the anomaly is seen at different energies
depending upon the redshift of the source. This in turn is not expected within the LIV scheme which predicts a fixed energy above which the optical depth
is suppressed. In principle, in models of D-branes, different line-of-sights could be affected in different ways \cite{2010PhLB..694...61E}. However, it would require a very unlikely
finetuning of the model in order to explain the observed effect. 
\item \textbf{Axion-like particles:}
 A more likely scenario is the mixing of the photon with a spin-0 boson which
would suppress the effect of pair production in the observed way. In the most
general form, mixing during propagation in the intergalactic medium with a
given domain size \cite{2011arXiv1106.1132D, 2009JCAP...12..004M} would have to
be combined with mixing in the source \cite{2009PhRvD..79l3511S} and the
Galactic magnetic field \cite{2008PhRvD..77f3001S}.  Given the large range of
possible combinations of parameters (domain size and product of magnetic field
and coupling $g_{a\gamma}$) including stochastic variations along the different line of
sights as well as an unknown shape of the true EBL, it is well beyond the scope
of this paper and the data available to derive a reliable estimate of the most likely coupling and
mass of the mixing partner of the photon.  However, a qualitative estimate comparison has been done. The result is shown in
Fig.~\ref{figtau2} where the same data are shown as in Fig.~\ref{fig4}. Overlaid is now a set of curves which
provide an estimate on how much the transparency would change in the presence of a pseudo-scalar field which couples to 
photons with $g_{a\gamma}=10^{-11}\mathrm{GeV}^{-1}$ in the presence of an intergalactic magnetic field of $1~\mathrm{nG}$
and domains of random orientation of the $B$-field and length of 5~Mpc. 
The chosen value for the field strength is not ruled out by 
observational bounds (see e.g. \cite{2009PhRvD..80l3012N} for
a compilation of available limits) \footnote{Recently, observational evidence
has been discussed that 
the intergalactic medium is efficiently heated through generation
of plasma-instabilities by powerful blazars \cite{2011arXiv1106.5494B,2011arXiv1106.5505P,2011arXiv1106.5504C,2011arXiv1107.3837P}. If this heating
mechanism is at work, it would imply a model-dependent 
upper limit on the field strength of $\approx 10^{-12}$~G.}.
 The average transfer matrix has been calculated using the
solution obtained in \cite{2009JCAP...12..004M} using self-consistently the same lower limit EBL. As can be readily seen from the graph,
the qualitative behaviour matches the observational data quite well.
\end{itemize}

\section{Summary and conclusion}
In the past years,  the VHE spectroscopy of extragalactic objects has been considerably extended in energy and redshift covered. 
The currently available data have been collected and
analysed for the first time in a comprehensive and consistent manner. The intrinsic spectral shapes have been recovered under the assumption of a minimum (guaranteed)
absorption of VHE photons in pair production processes. 
A simple unbinned test has been introduced to search for the emergence of spectral features  at the transition from optically thin to optically thick: the scatter
of the spectral measurements around the extrapolation of a fit-function to the optically thin part of the spectrum ($\tau <1$) 
is compared between well-defined samples covering ranges in optical depth of $1\le \tau < 2$
and $2\le \tau$. The two samples  show differences at a significance level of $S=4.2~\sigma$ 
({$S=5.6~\sigma$ using the additional assumption of gaussianity of the residuals verified
with the data}) indicating that the observed absorption is smaller than
the minimum absorption assumed. A number of systematic effects (e.g. shift of energy scale, flux bias at the end of the spectra) have been considered but found
unlikely to provide the exclusive explanation for the observed effect. 
Source intrinsic features are unlikely  to explain the upturn of the spectra at $\tau>2$
unless an unnatural finetuning  of the source with the optical depth at which it is observed exists. 
 As a result of the study presented here, we conclude that  the observations indicate
the presence of a suppression of pair production during the propagation of VHE photons which is coined ``pair-production anomaly'' (PPA). 
 The data do not allow to constrain the properties of the PPA but a plausible explanation is provided by coupling photons to a pseudo-scalar
(axion-like particles) via intergalactic magnetic fields.
\bibliographystyle{JHEP}
\acknowledgments
We thank Tanja Kneiske, Martin Raue, Andrei Lobanov, Javier Redondo, Alessandro Mirizzi,  Andreas Ringwald, and Marco Roncadelli  
for valuable input and discussions
on this theme. We thank the anonymous referee for constructive comments.
MM acknowledges the support of the Hamburg cluster of excellence \textit{Connecting Particles with the Cosmos}. 
DH acknowledges the support of the collaborative research center SFB~676 \textit{Particles, Strings, and the Early Universe}. 
This research has made use of NASA's Astrophysics Data System.
\bibliography{mybib}
\appendix
\section{Energy spectra}
{ The figures \ref{sf01}-\ref{sf07} show the energy spectra of the
seven sources with spectral measurements in the optically thin as
well as in the optically thick regime.}
\begin{figure}
\subfloat[]{\label{sf01}\includegraphics[width=0.5\linewidth]{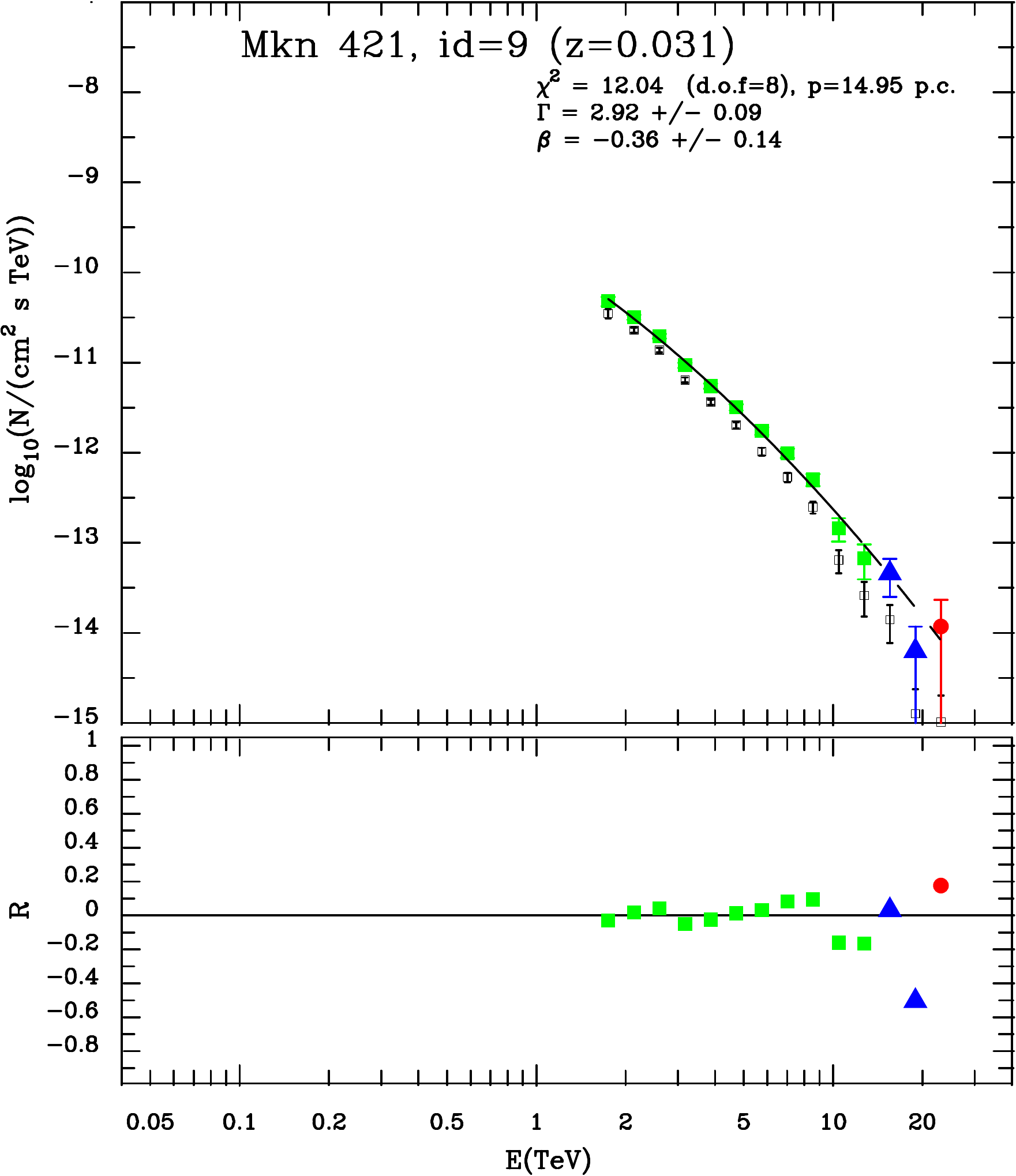}}
\subfloat[]{\label{sf02}\includegraphics[width=0.5\linewidth]{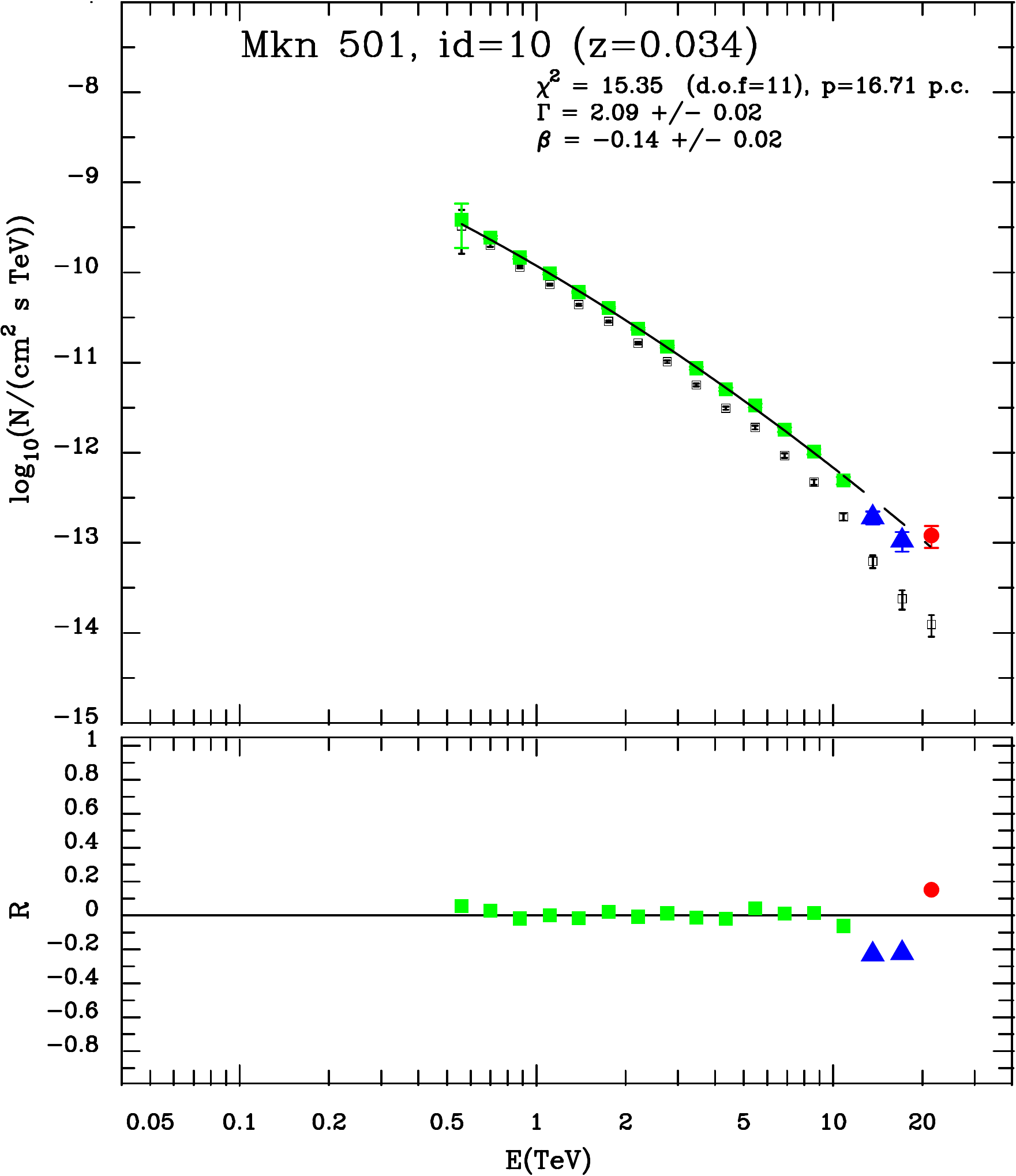}}
\\
\subfloat[]{\label{sf03}\includegraphics[width=0.5\linewidth]{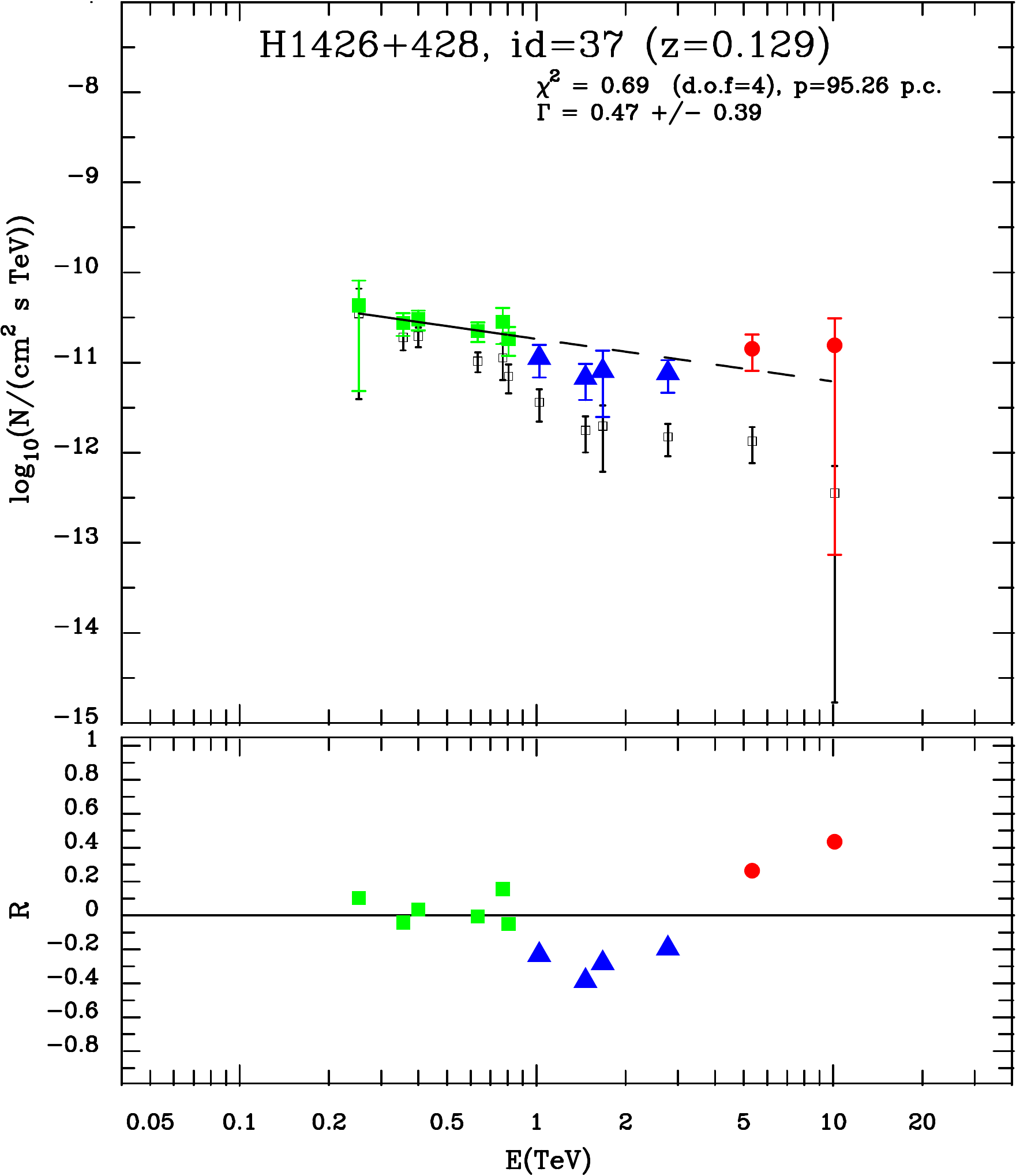}}
\subfloat[]{\label{sf04}\includegraphics[width=0.5\linewidth]{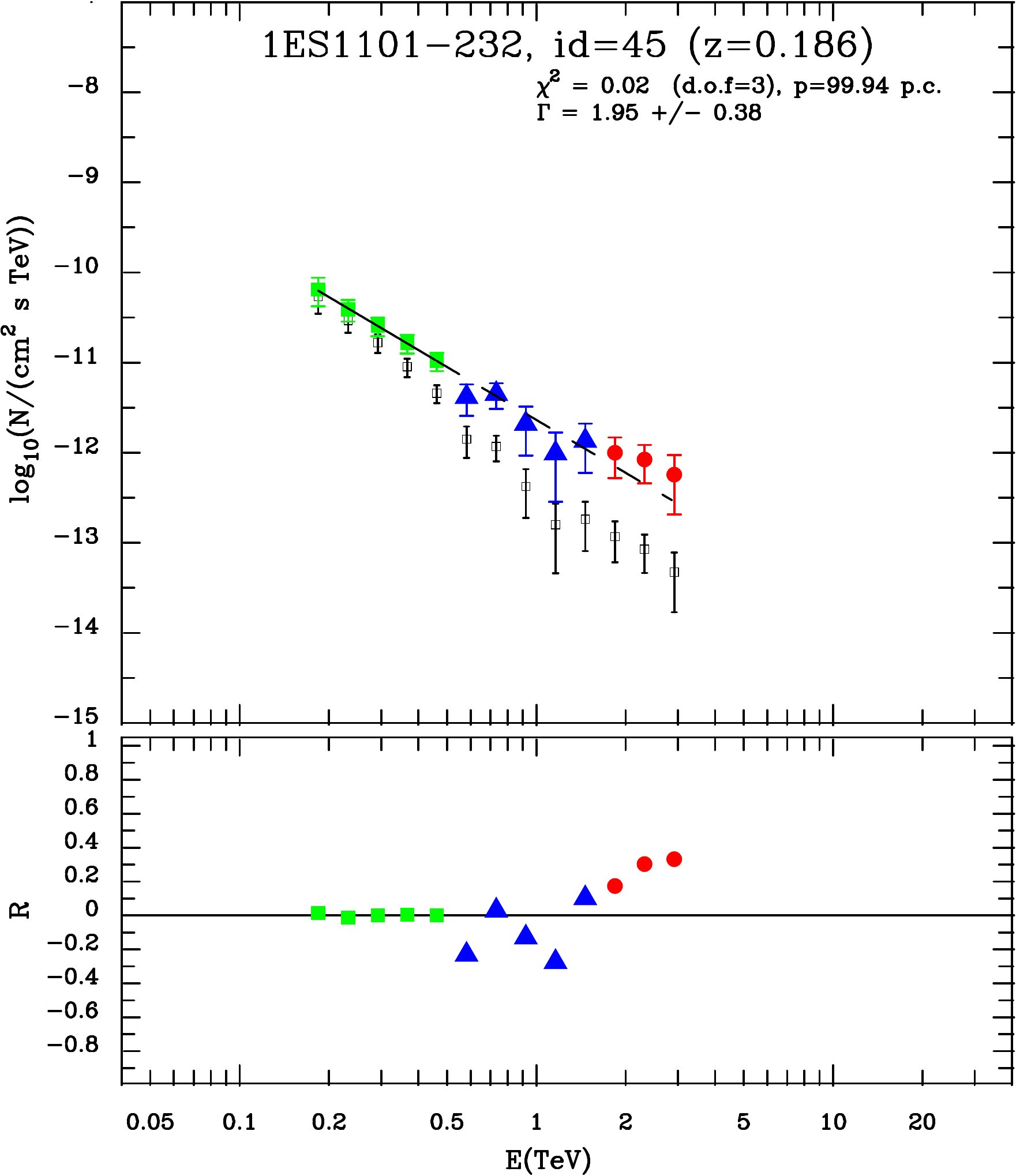}}

\caption{VHE energy spectra of AGN: The black open squares indicate the measured differential spectra while the coloured 
solid markers are the measurements corrected for absorption (squares/green: $\tau<1$, triangles/blue: $1\le\tau<2$, bullets/red: $2\le \tau$).
\label{fig:spectra}}
\end{figure}
\begin{figure}
\ContinuedFloat
\subfloat[]{\label{sf05}\includegraphics[width=0.5\linewidth]{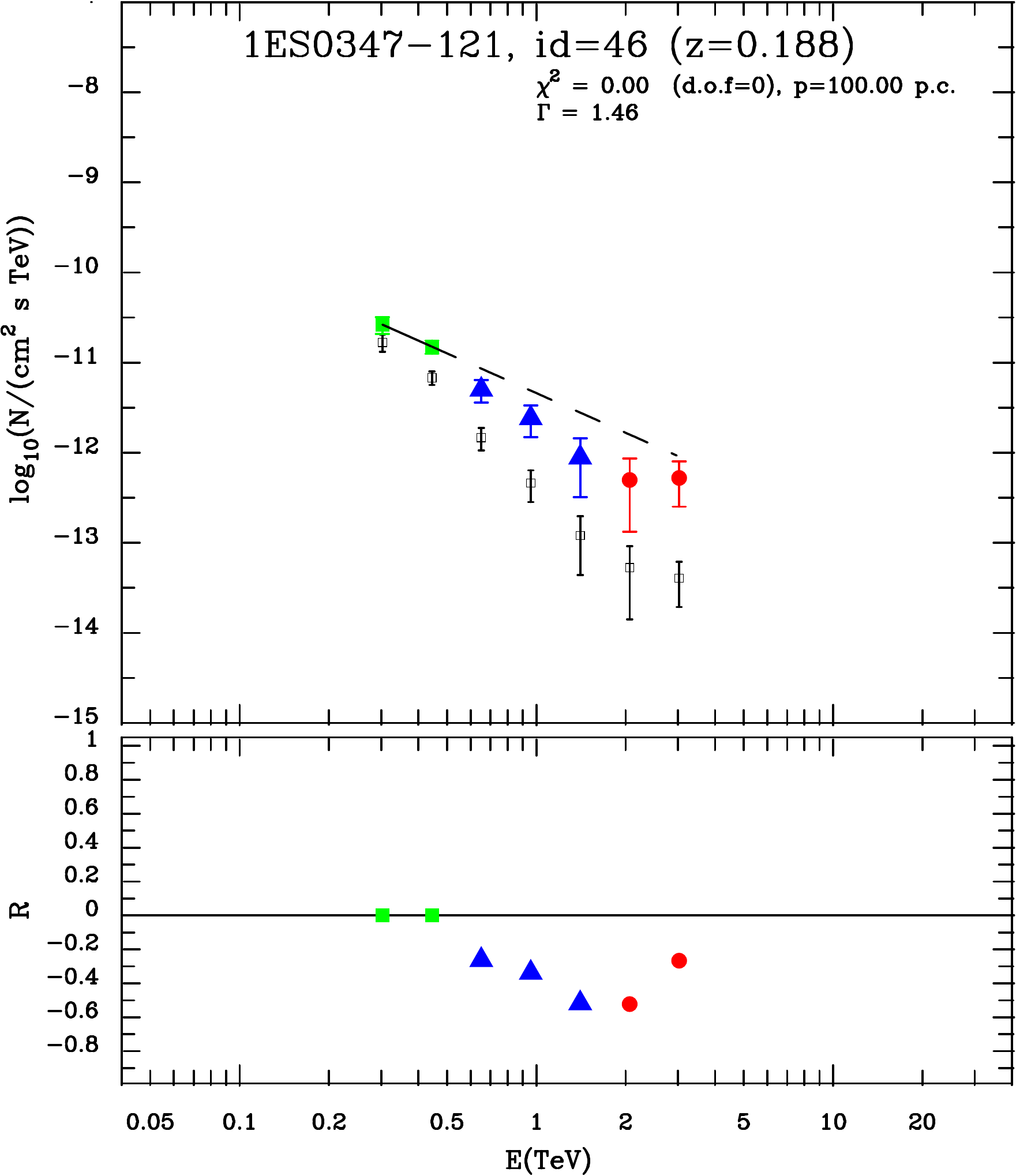}}
\subfloat[]{\label{sf06}\includegraphics[width=0.5\linewidth]{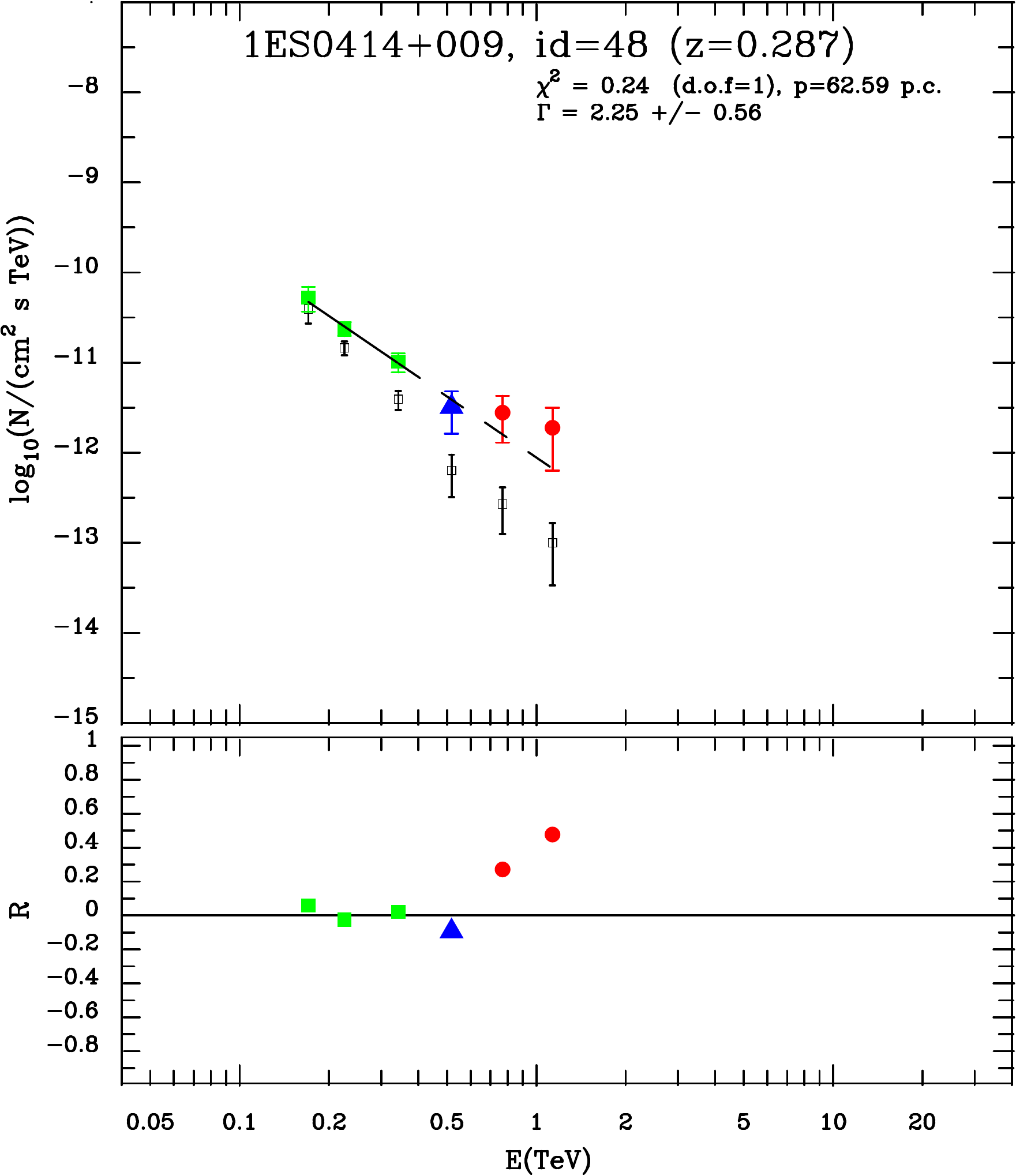}}
\\
\subfloat[]{\label{sf07}\includegraphics[width=0.5\linewidth]{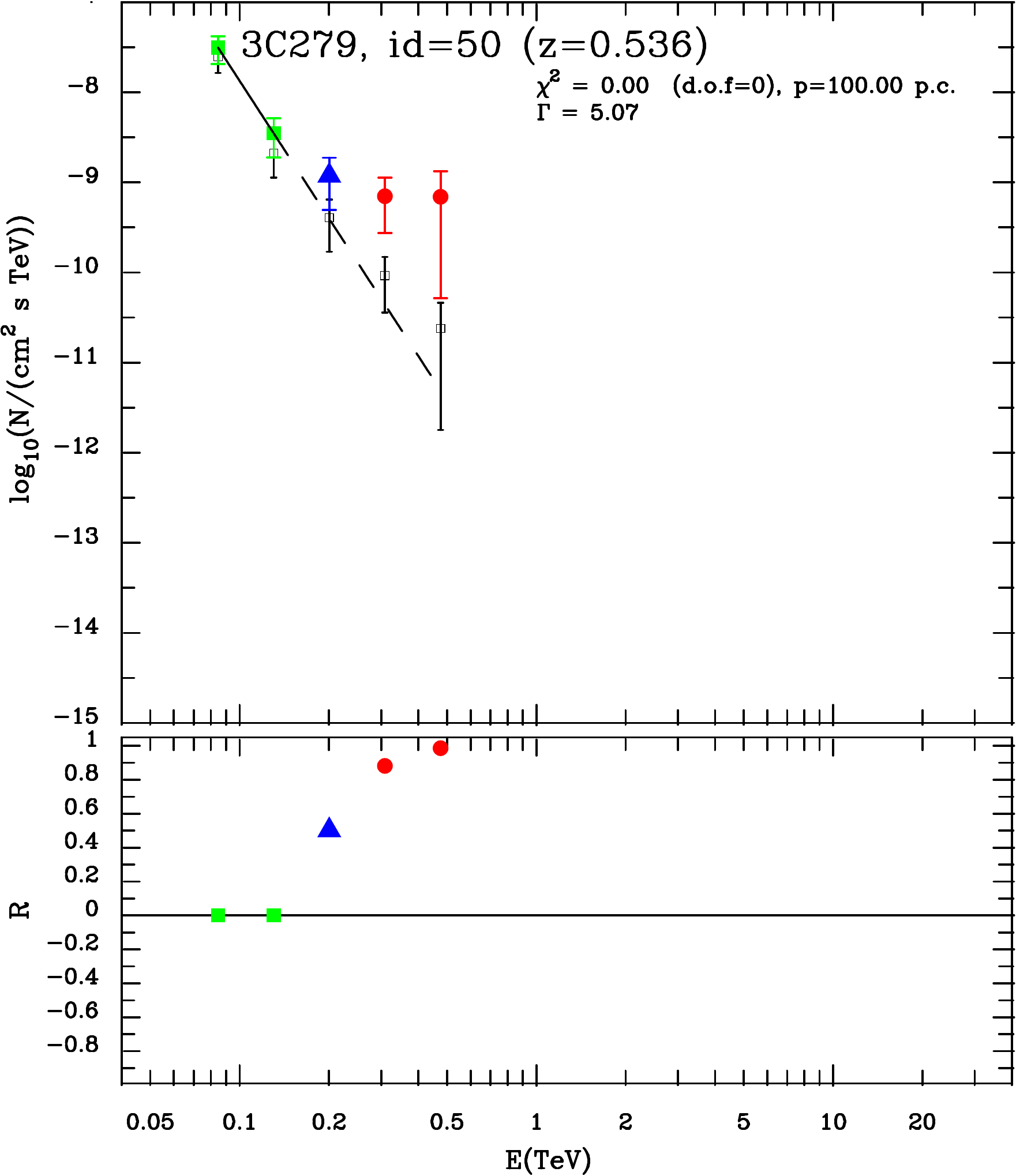}}
\caption{ Energy spectra of Blazars: The black open squares indicate the measured differential spectra while the coloured 
solid markers are the measurements corrected for absorption (squares/green: $\tau<1$, triangles/blue: $1\le\tau<2$, bullets/red: $2\le tau$).
\label{fig:spectra}}
\end{figure}
\section{Distribution of averaged residuals}
{ The test introduced in Section~2 is based upon the null hypothesis that
the independent 
samples of $R(\Phi_i\in \mathcal{B})$ and $R(\Phi_i\in \mathcal{S})$
stem from the same underlying unknown distribution. 
The best-fit parameters
were derived by a $\chi^2$-minimization to the optically thin data-points.
In the following, this assumption is tested by investigating
the behaviour of the residuals $\chi_i$ as defined in Eqn.~2.2  after fitting
$f_\mathrm{id}$  to the entire spectrum. In Fig.~\ref{app1}, the
normalised histogram of residuals is shown in
three exclusive intervals of energy together with 
a normal distribution with $\mu$ and $\sigma$ chosen to match the
mean and $\sqrt{var}$ of the samples chosen. The samples are consistent
with normal distributions: Anderson-Darling tests estimates probabilities of
$0.77$, $0.68$, and $0.43$ for the three samples to be normal distributed.
The average values of the residuals are all within one standard deviation 
of the mean consistent with zero. The scatter of the samples is 
distinctly different from the expected value, indicating that the experimental errors are over-estimated. \\
When taking the same residuals and considering the samples split according
to the optical depth (using the same ranges as for the previous tests), 
again fairly normal distributed residuals are found (Anderson-Darling test:
$0.07$, $0.44$, and $0.12$, the width is again too narrow). The average residuals 
in the first two bins of $\tau$ are consistent with zero.
For the last bin, the average residual of $\mu=0.73\pm0.13$ is
significantly different from zero ($S=5.6~\sigma$), confirming
the result obtained using the Kolmogorov-Smirnov test on the
samples of $R$. 
The average values are compared  for the different samples in Figs.~\ref{app3}-\ref{app4}.

\begin{figure}
  \subfloat[]{\label{app1}\includegraphics[width=0.45\linewidth]{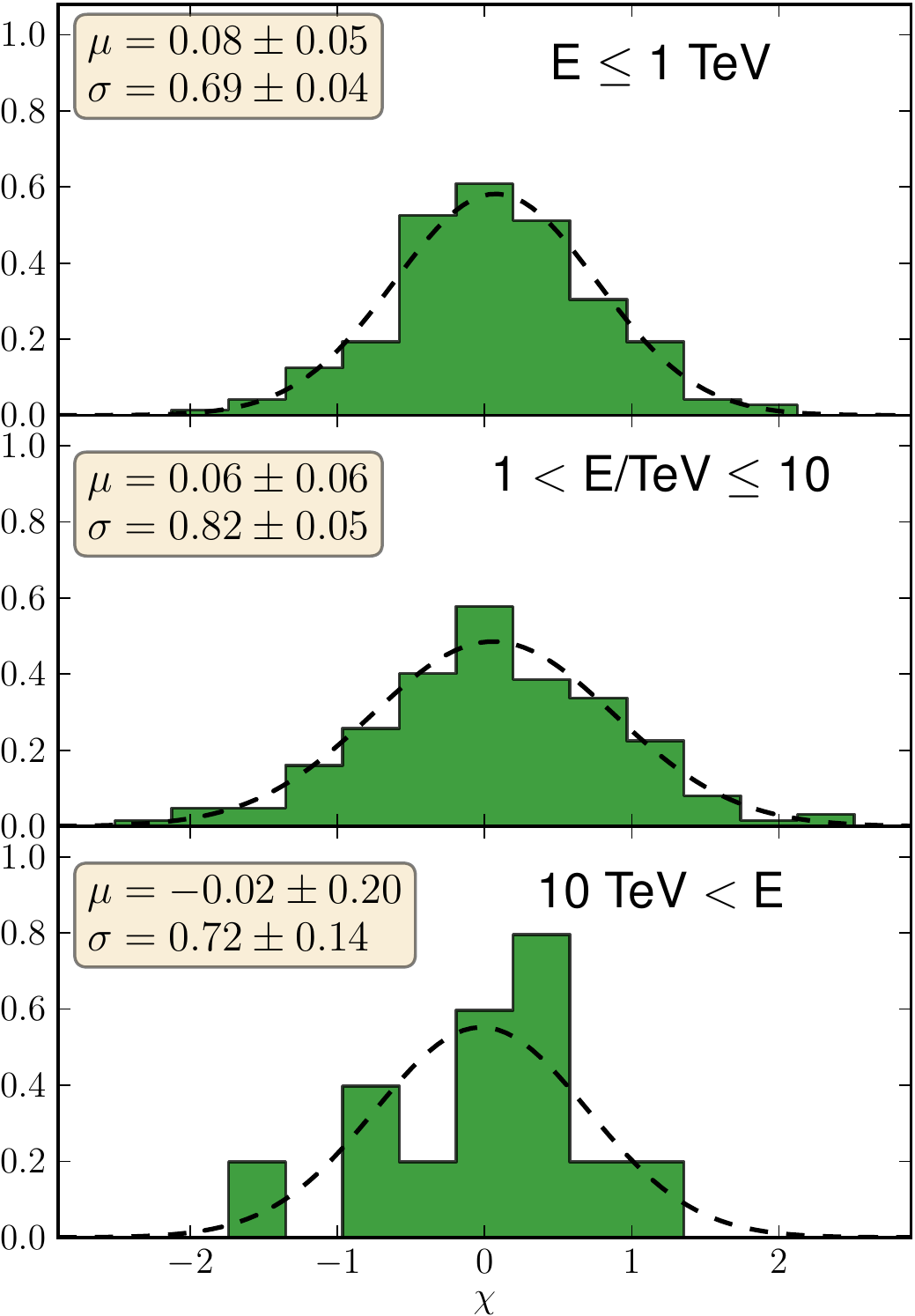}}
\hspace*{0.05\linewidth}
  \subfloat[]{\label{app2}\includegraphics[width=0.45\linewidth]{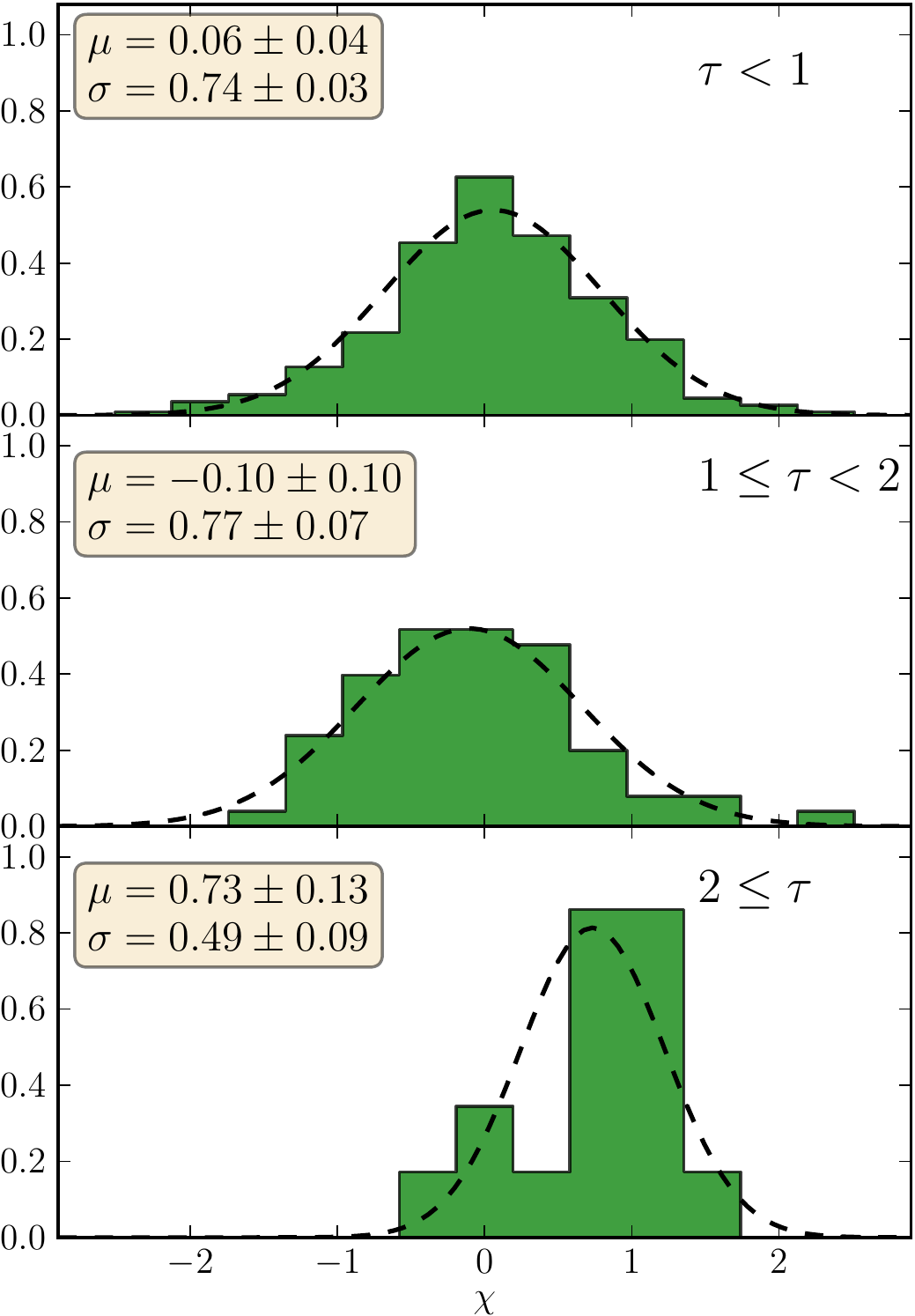}}
\caption{Normalised histograms of residuals in three intervals of
energy (a) 
and optical depth (b),
overlaid with dashed lines are the best-fit normal distributions.}
\end{figure}
\begin{figure}
  \subfloat[]{\label{app3}\includegraphics[width=0.5\linewidth]{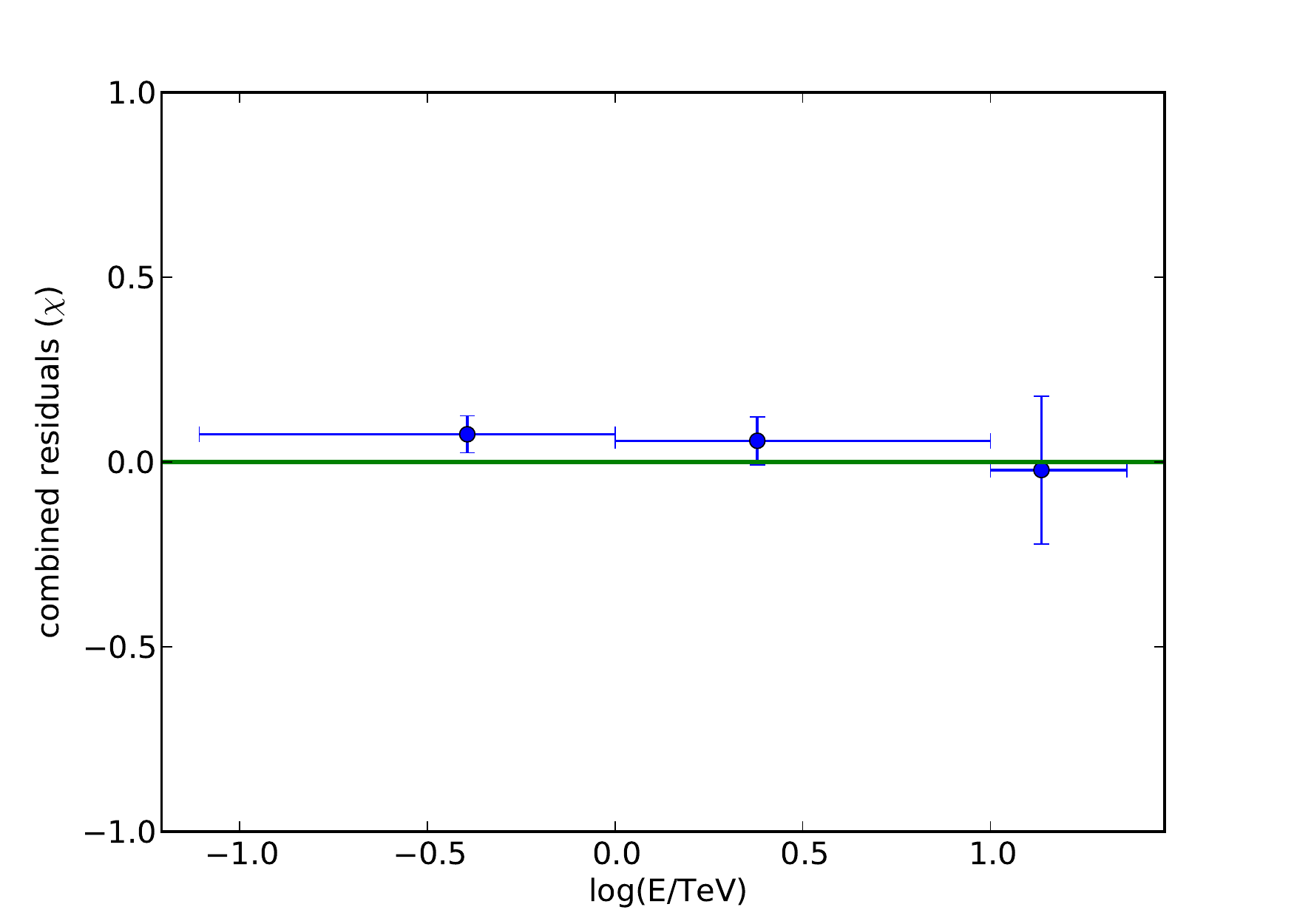}}
  \subfloat[]{\label{app4}\includegraphics[width=0.5\linewidth]{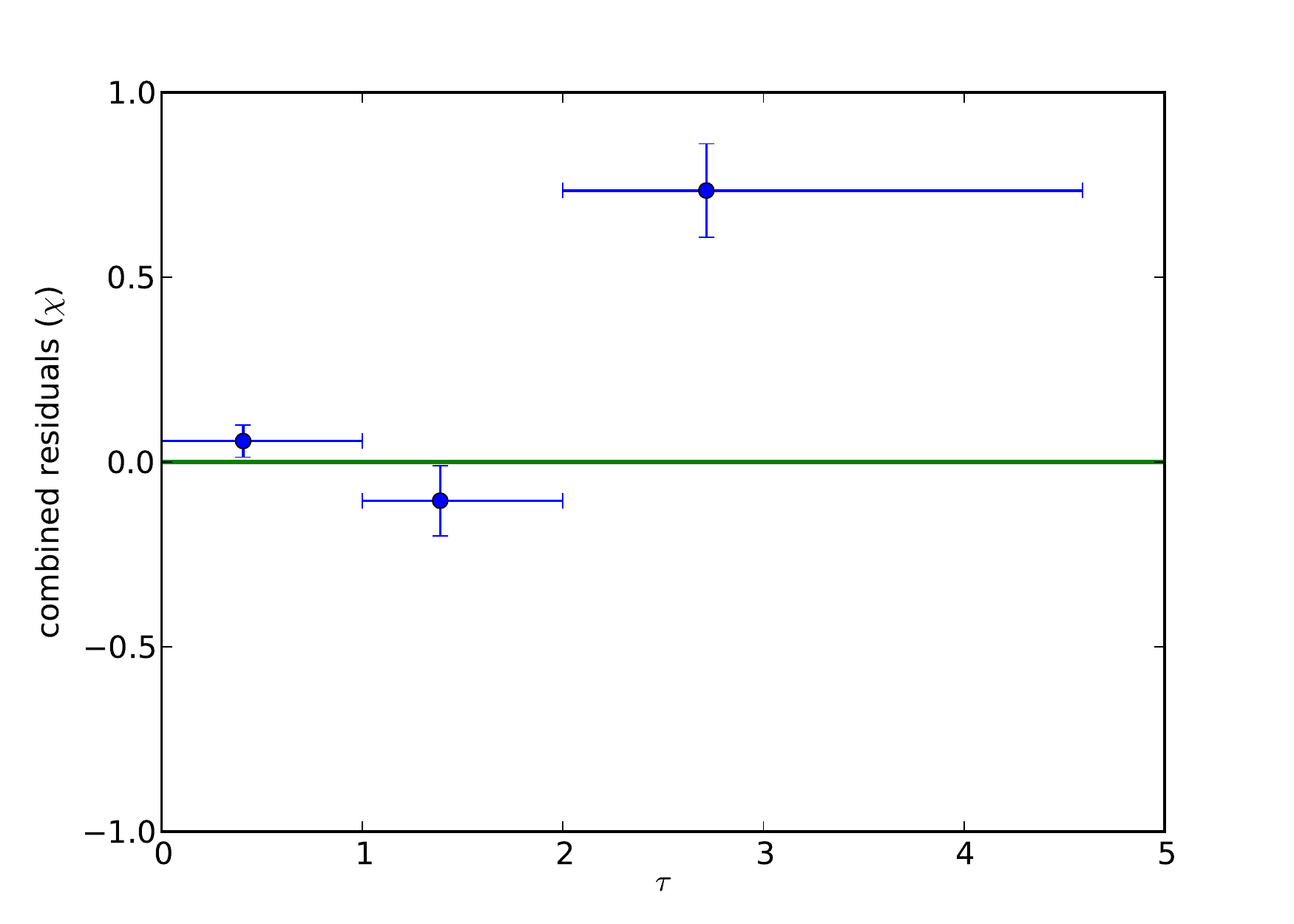}}
\caption{Average residuals ($\mu$) and errors for three intervals of energy 
(a) 
and optical depth (b), 
the horizontal error bars indicate the range of values included in 
each bin, the vertical error bars show the $1~\sigma$ uncertainty on $\mu$.}
\end{figure}

\end{document}